\documentclass[]{spie}  

\usepackage{amsmath,amsfonts,amssymb}
\usepackage{graphicx}
\usepackage[colorlinks=true, allcolors=blue]{hyperref}
\usepackage[normalem]{ulem}

\title{Athena charged particle diverter simulations: effects of micro-roughness on proton scattering using Geant4}

\author[a]{Jean-Paul Breuer}
\author[a,b]{G\'abor Galg\'oczi}
\author[c]{Valentina Fioretti}
\author[d]{Jakub Zl\'amal}
\author[d]{Petr Li\v{s}ka}
\author[a]{Norbert Werner}
\author[e]{Giovanni Santin}
\author[e]{Nathalie Boudin}
\author[e]{Ivo Ferreira}
\author[e]{Matteo Guainazzi}
\author[f]{Andreas von Kienlin}
\author[g]{Simone Lotti}
\author[i]{Teresa Mineo}
\author[g]{Silvano Molendi}
\author[j]{Emanuele Perinati}
\affil[a]{Department of Theoretical Physics and Astrophysics, Faculty of Science, Masaryk University, Brno, Czech Republic}
\affil[b]{Institute of Physics, E\"otv\"os Lor\'and University, Budapest, Hungary}
\affil[c]{INAF/Osservatorio di Astrofisica e Scienza dello Spazio, Bologna, Italy}
\affil[d]{CEITEC, Brno University of Technology, Brno, Czech Republic}
\affil[e]{ESTEC/ESA, Keplerlaan 1, 2201AZ Noordwijk, The Netherlands}
\affil[f]{Max Planck Institut f\"ur extraterrestrische Physik, Garching bei M\"unchen, Germany}
\affil[g]{INAF/Istituto di Astrofisica Spaziale e Fisica cosmica, Milan, Italy}
\affil[i]{INAF/Istituto di Astrofisica Spaziale e Fisica Cosmica, Palermo, Italy}
\affil[j]{Eberhard Karls University, T\"ubingen, Germany}

\authorinfo{Further author information: (Send correspondence to Breuer \& Galg\'oczi)\\Breuer: E-mail: jeanpaul.breuer@gmail.com\\  Galg\'oczi: E-mail: gaborgalgoczi@gmail.com}

\begin{document} 
\maketitle

\begin{abstract}
The last generation of X-ray focusing telescopes operating outside the Earth's radiation belt discovered that optics were able to focus not only astrophysical X-ray photons, but also low-energy heliophysical protons entering the Field of View (FOV). This ``soft proton'' contamination affects around 40\% of the observation time of XMM-Newton. The ATHENA Charged Particle Diverter (CPD) was designed to use magnetic fields to move these soft protons away from the FOV of the detectors, separating the background-contributing ions in the focused beam from the photons of interest. These magnetically deflected protons can hit other parts of the payload and scatter back to the focal plane instruments. Evaluating the impact of this secondary scattering with accurate simulations is essential for the CPD scientific assessment. However, while Geant4 simulations of grazing soft proton scattering on X-ray mirrors have been recently validated, the scattering on the unpolished surfaces of the payload (e.g. the baffle or the diverter itself) is still to be verified with experimental results. Moreover, the roughness structure can affect the energy and angle of the scattered protons, with a scattering efficiency depending on the specific target volume. Using Atomic Force Microscopy to take nanometer-scale surface roughness measurements from different materials and coating samples, we use Geant4 together with the CADMesh library to shoot protons at these very detailed surface roughness models to understand the effects of different material surface roughnesses, coatings, and compositions on proton energy deposition and scattering angles. We compare and validate the simulation results with laboratory experiments, and propose a framework for future proton scattering experiments.
\end{abstract}

\section{INTRODUCTION}
\label{sec:intro}  
The last generation of X-ray observatories were extremely successful in resolving astrophysical objects and answering many questions regarding the cosmos and our observable universe. Launched in 1999, the NASA Chandra X-ray Observatory (Chandra) and the European Space Agency (ESA) X-ray Multi-Mirror Mission (XMM) Newton telescopes are complementary missions, with Chandra providing better spatial resolution, while XMM-Newton provides a larger collecting area offering better spectroscopic information. With these modern observatories, it was discovered that low energy protons ($\le$300 keV) can also scatter through the optics at low angles and get funneled towards the focal plane. These ``soft protons'' manifest as sudden flares in the detector background count rates. They populate the interplanetary space, as well as different parts of the Earth's magnetosphere, and are collected during the orbit cycle of the observatory \cite{walsh,system1,system2,system3}.

This soft proton contamination is highly variable over time and can not be efficiently modeled, as their signal in the detector is registered as events similar to the ones created by X-ray photons. Current treatment methods involve looking for anomalous periods of high count rate in the detector light curves and removing all data during these events \cite{snow}. This soft proton contamination currently affects around 40\% of XMM-Newton observations, and can be a significant contribution to the background for future X-ray missions \cite{system2}. Over the last 20 years, many attempts have been made to better understand and predict soft proton flares \cite{snow,ml}. 


ESA's second large-class mission is also the next generation X-ray focusing observatory mission, the Advanced Telescope for High-ENergy Astrophysics (Athena), which will launch in the 2030's and will provide unprecedented spectroscopic and imaging capabilities from the 0.5 keV to 12 keV band \cite{athenawhite}. Athena will use Silicon Pore Optics (SPO) technology together with two focal plane instruments, the Wide Field Imager (WFI) and a calorimeter spectrometer, the X-ray Integral Field Unit (X-IFU), with the science goal of exploring the hot and energetic Universe. Athena should be placed in a halo orbit around L1, the first Lagrange point of the Sun-Earth system, after studies of the plasma environment in both L1 and L2 \cite{valaas,L1}.

To limit the soft proton contamination, the WFI has a challenging scientific requirement for the soft proton background to be $< 5\times 10^{-4}$ cts/cm$^2$/keV/s, less than 10\% of the non-focused background requirement \cite{wfibkg}. Athena will directly address this soft proton contamination problem using a new device, which will be placed between the optics and the detector systems. This Charged Particle Diverter (CPD) will be close to the instruments in the Science Instrument Module (SIM), and will create a magnetic field which will deflect the soft energy protons outside the field of view\cite{cpd}. However, simulations also predict a potential secondary scattering of the deflected protons with the CPD walls and surfaces of the focal plane assembly (e.g. WFI baffle). To test if the CPD is compliant with the background requirement, it is first necessary to understand how surface roughness and composition can affect the proton energy and angles after scattering. To do this, an accurate simulation of the proton scattering with the surfaces of the focal plane assembly as well as the wall of the WFI baffle and CPD must be performed to know the impact of the roughness on the scattering efficiency. Previous validations of Geant4 simulations of soft proton scattering were performed at grazing angles and on polished X-ray mirror samples \cite{valaas}. Since roughness and composition can affect the proton energy and angles after scattering, laboratory measurements of proton scattering from representative samples were taken to verify dedicated Geant4 and SRIM simulations. This work specifically focuses on the validation of the proton scattering simulations performed in Geant4 using the experimental results, with the preliminary results already giving some general implications on the roughness and composition effects on low-energy proton scattering. 

This work fits into the context of a larger scale experiment to characterize the soft proton background, as part of the scientific assessment of the CPD. Our working group aims to perform end-to-end simulations of these soft protons, starting with ray-tracing simulations of the proton interaction with the mirror assembly, followed by the effect of the CPD induced magnetic field on the way towards the detector systems, the proton interaction and scattering against the walls of the focal assembly, and finally the endpoint of these protons.

Section 2 aims to explain the surface data, the experimental set up, as well as the experimental set up within Geant4 and SRIM. Section 3 will discuss a statistical systematic uncertainty introduced as a result of a computational optimization originating from the binning of the surface roughness model. Sections 4 and 5 will discuss the results and validity of the simulations, and end with some concluding remarks and details for the future work.

\section{DATA EXPLORATION, EXPERIMENTAL SET-UP, AND GEANT4 SET-UP}\label{sec:data} 

\subsection{SURFACE IMAGING}
The basis for all of the experiments began with measurements of the WFI baffle surface, performed via Scanning Probe Microscope (SPM) Bruker Dimension Icon at Brno CEITEC Nano\footnote{\url{https://www.ceitec.eu/scanning-probe-microscope-bruker-dimension-icon/e1359}}, which gave ASCII text output of a 512$^2$ pixel matrix of measured values spanning the 92$^2$ $\mu$m$^2$ dimensions of the physical sample. The atomic composition of the sample was also measured, by EDS X-Max 20 by Oxford Instruments installed in the Scanning Electron Microsope TESCAN Mira in BUT CEITEC Nano\footnote{\url{https://nano.ceitec.cz/scanning-electron-microscope-e-beam-writer-tescan-mira3raith-lis-mira/}}. The WFI baffle sample has an average surface roughness of 6.5 $\mu$m, a calculated density of 2.15 g/cm$^3$, and also has an optical coating called ``Magic Black'', with the composition in weight of 44\% Oxygen and 56\% Aluminum.

\begin{figure} [ht]
   \begin{center}
   \begin{tabular}{c} 
   \includegraphics[width=1\textwidth]{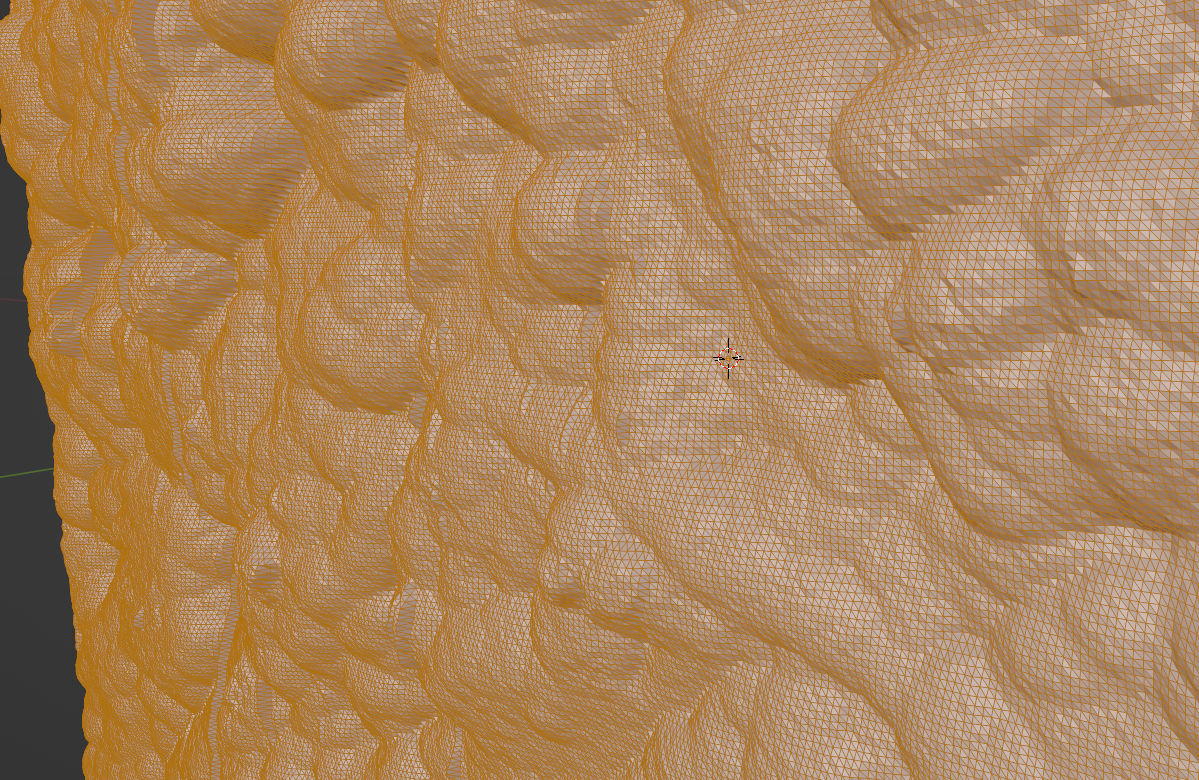}
	\end{tabular}
	\end{center}
   \caption[Example Surface]{Highly detailed CAD model of the WFI baffle surface sample, where over a half million triangles represent the 92$^2$ $\mu$m$^2$ area. The samples were imported into Geant4 using the CADMesh library and measured atomic composition, and were created from measurements using Matlab and Blender. \label{fig:sample}}
\end{figure}

The first step to prepare this measurement for the simulations is to convert the measured data from a text file into a Standard Tessellation Language (STL) file, the Computer Aided Design (CAD) standard file format that describes the whole geometry of an object in the form of a triangular mesh. Matlab has some user-defined functions that manage to make this conversion to STL by first defining the sample as a surface mesh object; however, this object has no thickness, and can be considered as an empty wireframe. The second step in creating the model is to import this empty wireframe STL file into Blender, a free and open-source 3D computer graphics software toolset, often used for 3D modelling. An example of this high resolution model is shown in Fig.~\ref{fig:sample}, where over a half million small triangles represent the 92$^2$ $\mu$m$^2$ surface. The 92$^2$ $\mu$m$^2$ model object is given a equivalent thickness of 300 $\mu$m. We then close and fill the extruded model, making it a full solid and not an empty shell, and rotate and flip the model so that it has the correct orientation after it is loaded into Geant4.

\subsection{EXPERIMENTAL SET UP}

\begin{figure} [ht]
   \begin{center}
   \begin{tabular}{c} 
   \includegraphics[width=\textwidth]{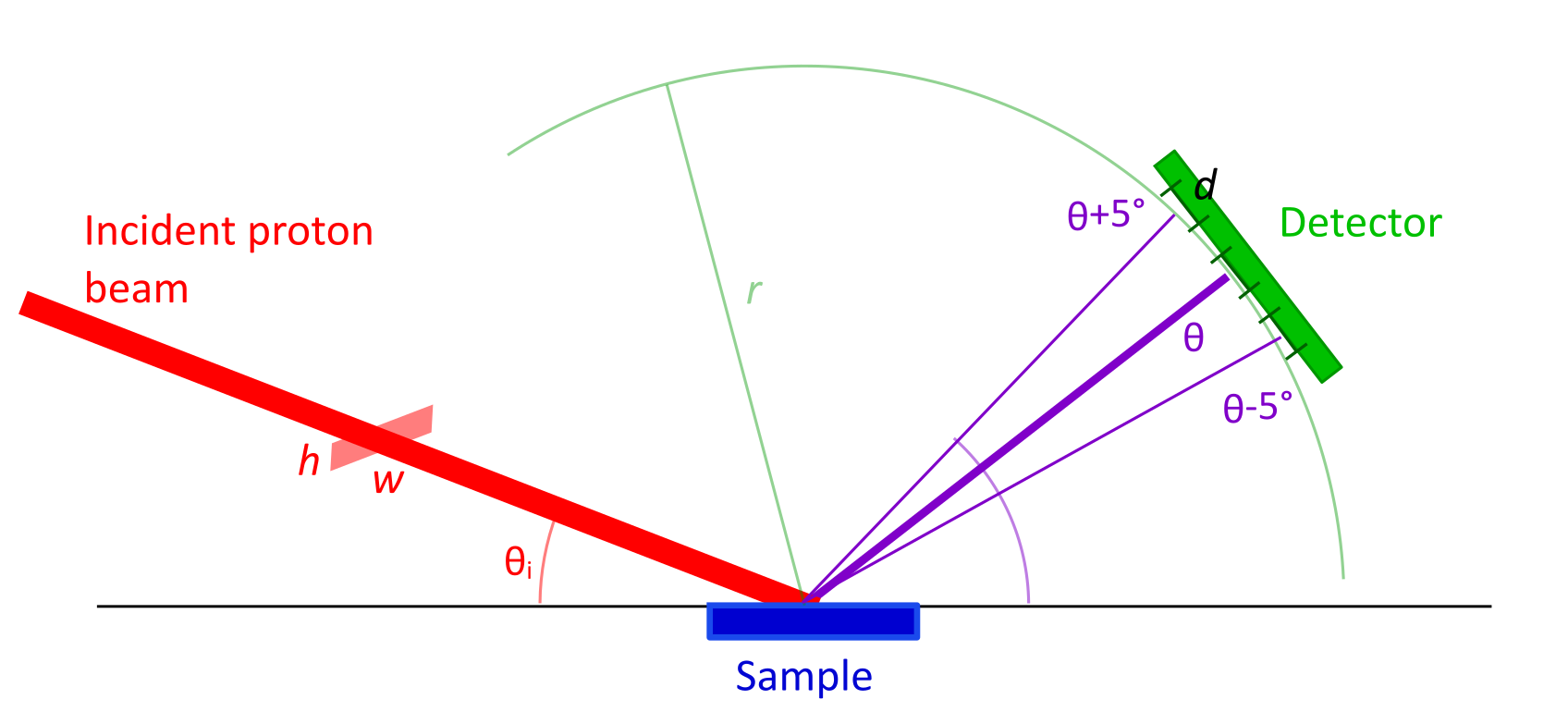}
	\end{tabular}
	\end{center}
   \caption[Proton Scattering Experiment]{Schematic view of the experimental setup: a particle beam shoots 100 keV protons at incidence angles of 20 degrees and 40 degrees, with the energy and scattering angles ($\theta$) of the deflected protons being measured by a detector at various positions, $\theta$ and $\theta \pm 5$. \label{fig:lab} }
\end{figure}

The WFI baffle sample was then placed in an experimental configuration as seen in Fig.~\ref{fig:lab}, used to characterize the scattering efficiency, performed at the Tandem laboratory of Uppsala University\footnote{\url{https://www.tandemlab.uu.se/?languageId=1}}. The WFI baffle surface sample was shot by a beam of 100 keV protons at an incidence angle of both 20 degrees and 40 degrees, with the energy and scattering angles ($\theta$) of the deflected protons being measured by a circular detector. The detector provides a signal from three separate areas, each with a diameter of 4 degrees, and simultaneously covers three different scattering angles per experiment, specifically, at scattering angles $\theta$ and $theta\pm5$ degrees. Fig.~\ref{fig:lab} shows a schematic of the experimental set up. The azimuth of proton scattering ($\phi$) is not explicitly measured with this experimental set up.

As the detector covers three scattering angles simultaneously, the detector was placed at 30 degrees, 50 degrees, and 65 degrees for the 20 degree incidence angle experiment, measuring all proton scattering angles $\theta$ from 30, 35, 40, etc. up to 70 degrees with respect to the measured surface plane.  For the 40 degree proton incidence angle case, the circular detector was moved to lower scattering angles, measuring scattered protons from 10, 15, 20, etc. up to 50 degrees. The experiment was performed several times to normalize the proton beam.

\subsection{GEANT4 MICRO-ROUGHNESS SIMULATION}
The Geant4 simulation framework is a toolkit library for particle transport codes based on Monte Carlo simulations, initially developed by CERN \cite{geant1,geant2,geant3}. All simulations in this work were based on Geant4 version 10.7.1, using the previously validated single Coulomb scattering (SS) model \cite{valaas}. To establish the framework to simulate the particle scattering, we additionally used version 1.1 of the CADMesh library, which allows us to load triangular mesh-based CAD files into Geant4\cite{cadmesh}. This combination of tools is particularly powerful, as it opens up many possibilities for future work and exploring the effects of different materials and treatments on scattering.

The Geant4 set-up is similar to the schematic from the experiment shown in Fig.~\ref{fig:lab}, where a particle gun shoots the digitized WFI baffle sample with a parallel beam of 100 keV protons at incidence angles of 20 and 40 degrees. The WFI baffle sample geometry is imported into Geant4 using the CADMesh library and is given the same atomic composition obtained from measurements. We deviate from the experimental set up in that we instead build a half-sphere around the sample surface to function as the main particle detector and record all the scattered particle hits. This half-sphere records the same proton energy and polar scattering angles $\theta$ as from the experiment, as well as the azimuthal scattering angles $\phi$. Additionally, this sphere provides a 2D angular distribution of $\theta$ vs $\phi$, which can be integrated over the detector area from the experiment, and provides the basis for validating the simulation results with the experimental results. 

We observed that shooting the 90 degree edges of the surface sample greatly increased the scattering efficiency of the protons at low scattering angles. We try to limit the particle interactions with the edges, sides, and corners of the CAD model, so the surface area of the reference sample which the proton gun covers is restricted to the central 72$^2$ $\mu$m$^2$ of the sample, corresponding to 80.4\% of the total sample. 



\subsection{SRIM SIMULATION}

A new algorithm for the calculation of the scattering of protons was implemented in a FORTRAN tracing code using SRIM \cite{srim}. SRIM itself calculates only ideal surface scattering, and in this case, was first used to pre-calculate the scattering of protons of various incidence energies and covering a full range of incidence angles. The total probability of proton scattering for each energy and incidence angle is then determined as the number of scattered protons to incident protons. 

Similar to the Geant4 and CADMesh implementation, the surface roughness of the WFI baffle sample is interpreted using 3D triangular meshes, but in this implementation, the protons do not penetrate or directly interact with the material. The FORTRAN code instead calculates the proton incidence angle with respect to the intersecting triangle and samples the derived probability distribution of scattering for that angle \cite{jakub}. The scattering direction of the proton is randomly chosen from the precalculated proton list, and the previous step is repeated in case of an intersection with another triangle from the 3D triangular mesh. The collecting of protons is done in FORTRAN, with detectors in the same positions according to the experiment, shown in Fig.~\ref{fig:lab}, making it possible to compare results with the experiment and other simulations.

\section{RESULTS}

\subsection{COMPARISON OF BINNED SAMPLES WITH FULL SAMPLES}

One of these SPM-measured surface roughness samples can have over 1 million small triangles, meaning that the memory and computational requirements for simulating protons are non-trivial. As a benchmark, one hundred thousand 100 keV protons at 40 degree incidence angle takes over 70 hours to simulate on the full resolution sample using 48 threads. To have reasonable statistics for validating this simulation via comparison with the laboratory measurements, we require several million proton simulations, equivalent to about a month of non-interrupted computation time.

The computational time requirements are reduced for both lower incidence angles and for samples with less surface roughness. This is because the scattering efficiency at lower incidence angles is higher, so obtaining better statistics requires less proton simulations; meanwhile, surfaces that are more smooth reduce the likelihood that a particle enters and exits the surface material several times before scattering. 

With the future objective of creating a proton transfer function and understanding the proton scattering efficiency at many different incidence angles and energies, the computational time requirement can go up to several years, and can easily increase to over a lifetime to further explore different levels of surface roughness, compositions, and surface treatments.

The memory specific requirements are mainly dependent on the requested binning of the angle in the 2D angular distribution and the total number of threads. Besides the basic memory requirements for loading the sample and running Geant4, the estimated memory requirements can be described by the cumulative product between the bit-size of double variables, the number of threads used in a simulation, the specified binning for the output histogram, and the number of protons simulated. Assuming histogram bin sizes of 1 deg$^2$, memory requirements are easily met even for the full resolution sample.

\begin{figure} [ht]
   \begin{center}
   \begin{tabular}{c} 
   \includegraphics[width=1\textwidth]{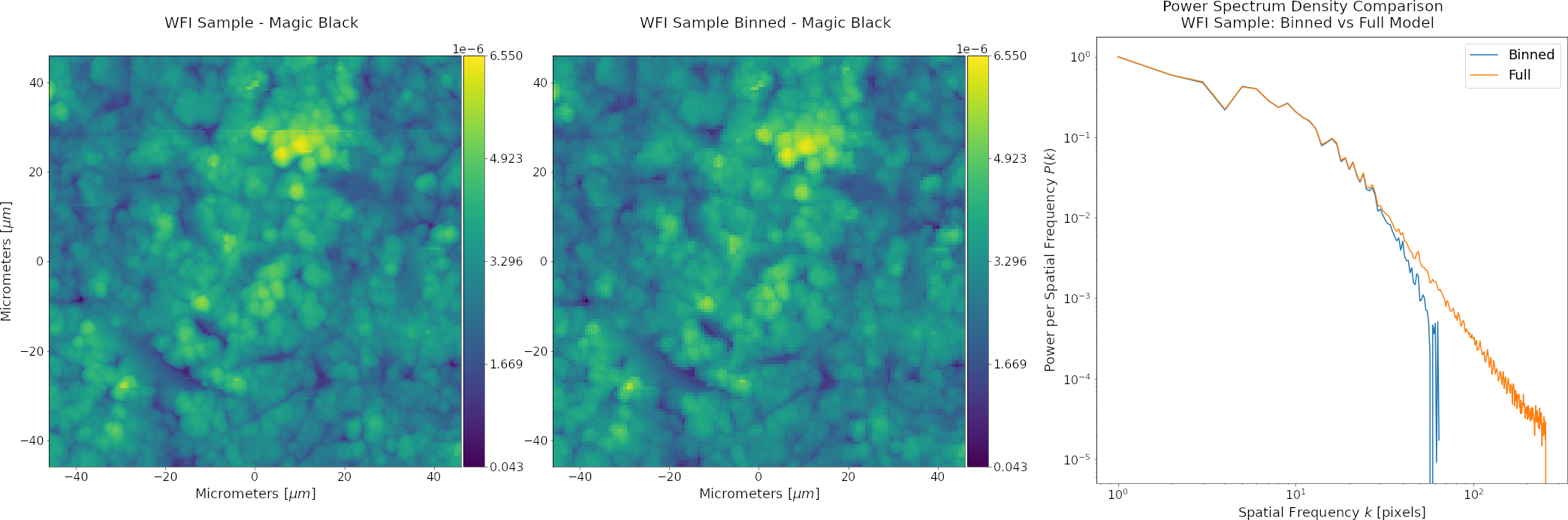}
	\end{tabular}
	\end{center}
   \caption[WFI Binned vs Full Model Comparison]{(\textit{left}) 2D visualization of the full WFI baffle sample roughness profile (with height in meters), treated with the ``Magic Black'' optical surface coating, same as the 3D version shown in Fig.~\ref{fig:sample}. (\textit{middle}) Visualization of the same sample, but binned by 4. (\textit{right}) Power spectrum density plot comparing the binned and full resolution samples with regards to the effective power of each spatial frequency, from small scales up to the Nyquist frequency. As expected, at large scales the sample looks fairly identical, but some information is lost on the small scales as it becomes averaged by the binning. This binning by 4 specifically reduces the effective power of smaller scale features by around 15\%. \label{fig:wfi_comp} }
\end{figure}

Binning the surface roughness samples was one of the optimization paths taken to reduce computation time. Extensive Geant4 simulations were run on both the full sample and the binned sample to determine the proton scattering efficiency of both distributions. Fig.~\ref{fig:wfi_comp} shows the full resolution WFI baffle sample, the binned version of the same sample, and a comparison of the two using a normalized power spectrum density plot. As evident in the plot, at least when binning the sample by four, 25\% of the smallest scale information is encoded in the averaging of the bins, but this binning only reduces the effective power of smaller scale features by around 15\%; in other words, the majority of the information is conserved.

\begin{figure} [ht]
   \begin{center}
   \begin{tabular}{c} 
   \includegraphics[width=1\textwidth]{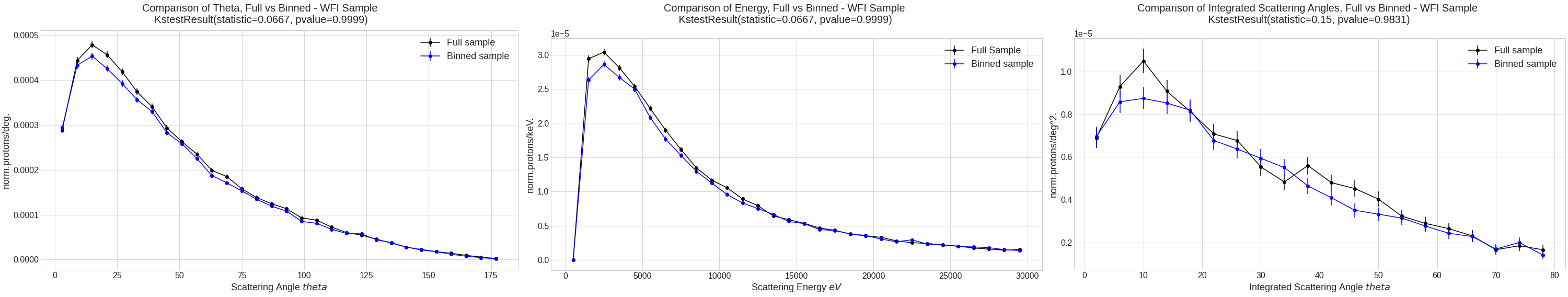}
	\end{tabular}
	\end{center}
   \caption[WFI Binned vs Full Model Results Comparison]{(\textit{left}) Comparison of proton scattering efficiency as a function of the scattering angle $\theta$ between the full resolution sample and the binned sample. (\textit{middle}) Comparison of proton scattering efficiency as a function of the proton energy between the full resolution sample and the binned sample. (\textit{right}) Comparison of the integrated proton scattering angles between the full resolution sample and the binned sample, in the frame of the laboratory detector. The pvalue result of the KS tests, from left to right, are 0.9999, 0.9999, and 0.9831, meaning that the difference between the full resolution and binned samples is less than 2\% when integrated to the frame of the lab experiment. \label{fig:binnedvs} }
\end{figure}

To test the effect of this binning on the proton scattering efficiency, millions of protons are simulated for both the full resolution WFI baffle sample, as well as the binned sample, over several weeks. The respective probability density distributions are then compared with each other using a Kolmogorov–Smirnov (KS) test to determine how similar the results are to each other, as well as to verify whether the binned samples are statistically self-similar enough to their own full resolution versions to be realistically used for the remainder of the simulation campaign. As can be seen in the left and middle of Fig.~\ref{fig:binnedvs}, the binning has a trivial effect in the shape of the distributions, with 99\% self similarity in proton scattering energy and scattering angle $\theta$. Meanwhile, the KS test determined that the final proton scattering efficiencies are 98\% percent similar to each other when integrating the results in the 2D angular distribution over the area of the detector used for the experiment, shown in the right plot of Fig.~\ref{fig:binnedvs}.

As mentioned previously, the benchmark of one hundred thousand 100 keV protons at 40 degree incidence angle took around 68 hours to simulate on the full resolution sample using 48 threads. The same configuration using the binned sample takes 6.3 hours for the same number of simulated protons. The difference in computation time between the full resolution sample and the binned sample is around 10x, so we can justify using the binned samples, since the statistical uncertainty introduced as a systematic of the binning is less than 2\% with respect to the integrated result. Millions of protons need to be simulated to obtain decent statistics, so this small change greatly reduces the computational overhead for the project at the cost of introducing a rather negligible systematic error.

\subsection{VALIDATION OF SIMULATIONS WITH EXPERIMENTAL RESULTS}

\begin{figure} [ht]
   \begin{center}
   \begin{tabular}{c} 
   \includegraphics[width=1\textwidth]{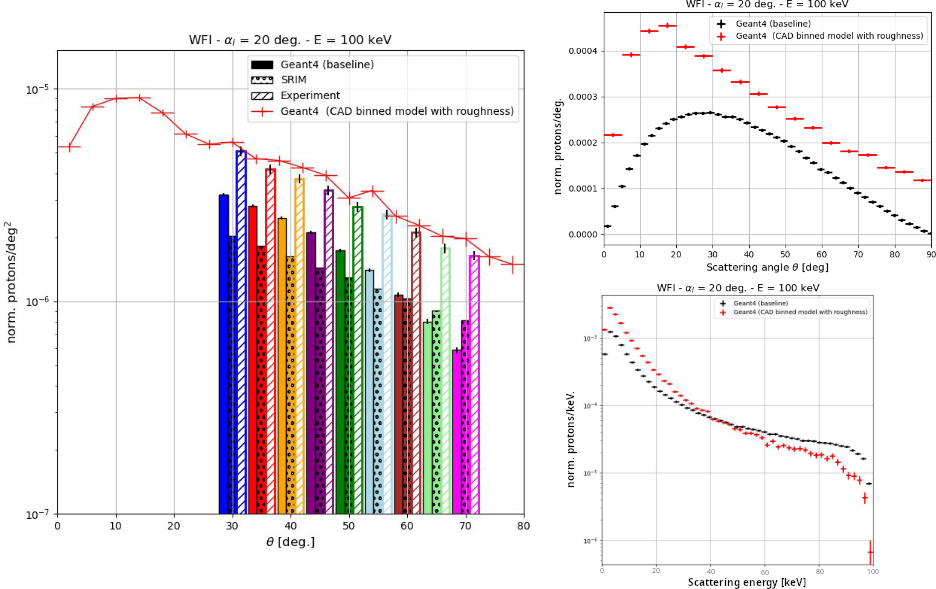}
	\end{tabular}
	\end{center}
   \caption[Validation 20 Deg]{From the CPD Scientific Assessment: Results for the WFI baffle sample with incidence angle of 20 deg, and 100 keV protons fired from the beam. (\textit{left}) Comparison between a baseline Geant4 simulation of an ideal surface, the SRIM simulation, the experimental results, and our Geant4 simulations with roughness and atomic composition. (\textit{top}) A distribution of the proton scattering angles after interactions with the WFI baffle surface sample, compared with the ideal surface. (\textit{bottom}) A distribution of the proton scattering energies after interactions with the WFI baffle surface sample, compared with the ideal surface. \label{fig:validation20} }
\end{figure}

\begin{figure} [ht]
   \begin{center}
   \begin{tabular}{c} 
   \includegraphics[width=1\textwidth]{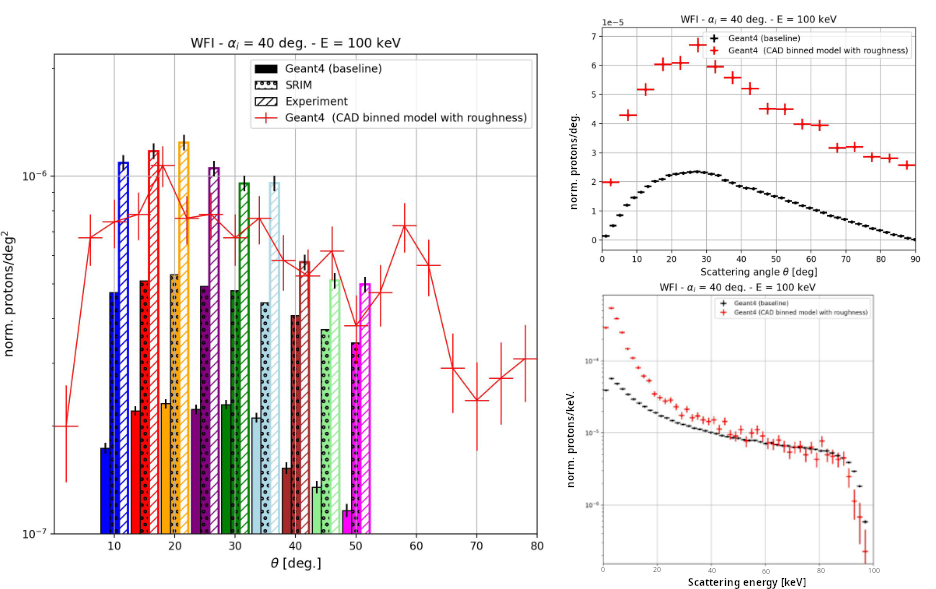}
	\end{tabular}
	\end{center}
   \caption[Validation 40 Deg]{From the CPD Scientific Assessment: Results for the WFI baffle sample with incidence angle of 40 deg, and 100 keV protons fired from the beam. (\textit{left}) Comparison between a baseline Geant4 simulation of an ideal surface, the SRIM simulation, the experimental results, and our Geant4 simulations with roughness and atomic composition. (\textit{top}) A distribution of the proton scattering angles after interactions with the WFI baffle surface sample, compared with the ideal surface. (\textit{bottom}) A distribution of the proton scattering energies after interactions with the WFI baffle surface sample, compared with the ideal surface. \label{fig:validation40} }
\end{figure}

Fig.~\ref{fig:validation20} and Fig.~\ref{fig:validation40} show the simulated proton scattering efficiencies compared with the measured results, with initial incidence angles of the proton beam at 20 degrees and 40 degrees, respectively. The left plot of Fig.~\ref{fig:validation20} and Fig.~\ref{fig:validation40} show our high resolution WFI baffle Geant4 simulations as a red line, with additional comparisons with other simulations, specifically, a simulation baseline from Geant4, where an ideal surface was used with no roughness, as well as results obtained from the SRIM simulations. The top and bottom right plots of Fig.~\ref{fig:validation20} and Fig.~\ref{fig:validation40} show, respectively, a distribution of the proton scattering angles and proton scattering energies, after interactions with the WFI baffle surface sample, compared with the ideal surface.

As can be seen in both Fig.~\ref{fig:validation20} and Fig.~\ref{fig:validation40}, our simulation results which include the binned roughness and exact composition of the samples are the closest with the experimental results. The SRIM simulation shows lower scattering efficiency than Geant4, but this could be due to an incorrect approximation of the scattering distribution, since many assumptions were necessary to optimize for computation time. The 40 degree proton incidence angle case is not a smoking gun, but is still the closest match to the experimental data out of all the simulations. Due to the lower scattering efficiency of larger incidence angles, more proton simulations and more computation time is required to reduce the size of the error bars and to better converge the integrated results to the true distribution. 

Looking at the top and bottom right plots of Fig.~\ref{fig:validation20} and Fig.~\ref{fig:validation40}, which compare the surface sample with roughness with a smooth ideal surface, we can see that modeling the roughness structure in Geant4 not only lowers the scattering angle of the protons, but also shifts the scattering proton energy, with less protons at higher energies and more protons below 40 keV. The shape of these two distributions also change a lot between the 20 degree to 40 degree cases, with the peak of the proton scattering angle shifting to larger scattering angles, and the same but more pronounced effect on the proton scattering energy distribution as previously observed.

\section{Conclusions and Future Activities} 
It is clear when comparing the simulation on a rough surface to the simulation on an ideal surface, that the current micro-roughness scale of 6.5 $\mu$m contributes to increase reflectivity at smaller scattering angles, with proton energies skewed to lower energies.

We also proved that including a refined modeling of the micro-structure of the surface where particles scatter improves the precision of the Geant4 simulation, with benefits in various applications where Coulomb scattering is the dominant interaction. Long computational requirements for simulations remain a concern, but we have been able to make some optimizations to this through the binning of surface sample CAD models with the minimal introduction of systematic uncertainties.

\begin{table}
\centering
\caption{Table of all of the available surface roughness samples measured by Scanning Probe Microscope (SPM) Bruker Dimension Icon. Base roughness indicates the arithmetical mean roughness in micrometers. All samples listed were simulated, but the sample investigated with the most computation and laboratory time, and focus of this report, is marked in bold.}
\label{tab:futuresamples}
\begin{tabular}{p{2cm}lp{8cm}c}
\hline
Material & Base Roughness & Surface Treatment & Sample No \\
\hline
 & Ra0.1 - polished & No (reference of flat surface) & 0 \\
AL7075 T7351 & Ra1.6 - machined & Surtec 650 (Finitec document ref. 1558830-914\_00) & 5 \\
 & Ra0.1 - polished & Solar black by Enbio & 10 \\
Ti6Al4V & Ra1.6 - machined & No & 19 \\
Kapton sample &  & Kapton foil on Al & Kapton \\
\textbf{WFI Baffle Sample} &  & \textbf{Magic Black} & \textbf{WFI} \\ 
\hline
\end{tabular}
\end{table}

The final goal of this research project is to create a proton transfer function to cut down on simulation time for other proton experiments. Now that the Geant4 simulation framework has been successfully validated by the experimental results, we can move on to the next phase of the project, which is to begin a simulation campaign for a greater range of proton incidence angles and energies.

\begin{table}
	\centering
	\caption{\label{tab:futurecomp} Summary of the samples with measured compositions and calculated densities. Atomic composition was measured by EDS X-Max 20 by Oxford Instruments installed in Scanning Electron Microscope TESCAN Mira in BUT CEITEC Nano. The sample investigated with the most computation and laboratory time, and the focus of this report, is marked in bold.}
	\begin{tabular}{lcccr}
	  \hline 
		Sample No & Composition & Density\\
        \hline 
        \textit{0} & 89.05\% Al, 2.5\% Mg, 2.5\% Cr, 5.5\% Zn, 1.8\% Cu, 0.5\% Fe, 0.4\% Si & 3.16 g/cm$^3$\\
		\textit{5} & 36.22\% Al, 41.33\% C, 16.25\% O, 1.23\% Zn, 4.96\% Zr & 2.55 g/cm$^3$\\
		\textit{10} & 2.73\% Al, 0.37\% Mg, 36.59\% C, 37.57\% O, 8.58\% P, 14.15\% Ca & 1.81 g/cm$^3$\\
        \textit{\textbf{WFI Baffle}} & \textbf{55.64.\% Al, 44.36\% O }& \textbf{2.15 g/cm$^3$} & \\
        \hline 
	\end{tabular}
\end{table}

At lower incidence angles, the incoming proton may enter and exit a rough surface sample several times before being scattered away, so we know that the scattering energy distribution is related to the final scattering angle to some extent; however, it is difficult to speculate on how the surface features directly contribute to these final distributions. By creating a probability density cube from all of our simulations, we can create an accurate sampling distribution function that statistically accounts for the effect of these unknown surface features, so given some input proton angle and energy we can quickly output a new scattering angle and energy for these protons.

Besides the WFI baffle sample presented in this report, several other materials and surfaces were measured and digitized, shown in Tab.~\ref{tab:futuresamples} and Tab.~\ref{tab:futurecomp}. Given that much infrastructure was developed for these high-performance computing routines, we hope to further explore these samples in the future and create various proton transfer functions for different materials and surfaces.

\acknowledgments 
JPB and NW are supported by the GACR grant 21-13491X. Part of the funding was provided by the Internal Grant Agency of Masaryk University, specifically the Operational Programme Research, Development and Education within the framework of the implemented project IGA MU reg. No. CZ.02.2.69/0.0/0.0/19\_073/0016943 as well as by the MASH3 grant. Computational resources were supplied by the project ``e-Infrastruktura CZ" (e-INFRA CZ LM2018140 ) supported by the Ministry of Education, Youth and Sports of the Czech Republic. The work was partly supported 
by the ESA go/esacloud computing infrastructure. CzechNanoLab project LM2018110 funded by MEYS CR is gratefully acknowledged for the financial support of the measurements at CEITEC Nano Research Infrastructure.

\bibliography{main} 
\bibliographystyle{spiebib} 

\end{document}